\documentclass{kluwer}
\newdisplay{guess}{Conjecture}
\begin{document}
\begin{article}
\begin{opening}
\tolerance=10000 \hbadness=10000
\def\verbfont{\small\tt} 
\def\note #1]{{\bf #1]}} 
\def\be{\begin{equation}}
\def\ee{\end{equation}} 
\def\bearr{\begin{eqnarray}} 
\def\eearr{\end{eqnarray}}
\def\barr{\begin{array}}
\title{VARIATION OF ACOUSTIC POWER WITH MAGNETIC FIELD AS SEEN IN GONG+ DATA}
\author{P. \surname{VENKATAKRISHNAN}}
\author{BRAJESH \surname{KUMAR}}
\author{S. C. \surname{TRIPATHY}}
\runningauthor{P. VENKATAKRISHNAN ET AL.}
\runningtitle{VARIATION OF ACOUSTIC POWER WITH MAGNETIC FIELD}
\institute{Udaipur Solar Observatory, Physical Research Laboratory, P.O. Box 198, 
Dewali, Off Bari Road, Udaipur 313 004, India\\
(E-mails: pvk@prl.ernet.in, brajesh@prl.ernet.in, sushant@prl.ernet.in)} 
\date{}
\begin{abstract}
The acoustic spectra in sunspots are known to be richer in higher frequency 
power. We have attempted a generalized study of the effect of magnetic fields 
on the shape of the acoustic spectrum using GONG+ bread-board 
data (spatial scale of $\sim$ 2 arc-sec per pixel) of 
11 May, 2000 and 12 June, 2000. The mean power spectra of the 
velocity oscillations were obtained by averaging over several spectra for 
different values of the magnetic field. With increasing magnetic field, 
the acoustic power 
increases at higher frequencies and decreases at lower frequencies with a 
transition at $\simeq$ 5 mHz. This behaviour is slightly different from 
earlier results obtained from SOHO/MDI data.
\end{abstract}

\end{opening}

\section{Introduction} 
The typical distribution of acoustic power of photospheric oscillations shows 
a maximum at around 5 min (Leighton, Noyes, and Simon, 1962) with a decrease 
to negligible power at higher frequencies. This behaviour of the acoustic 
spectrum has been understood in terms of trapped oscillations in a cavity. The 
eigen-functions of the natural modes of oscillations peak at different depths 
in the solar interior. The band of oscillations in the region of 5 minutes 
peak in the convection zone, which seems to be the dominant source of excitation 
of the solar oscillations (Goldreich, Murray, and Kumar, 1994). 
There have been some investigations on the behaviour of the acoustic spectrum 
in sunspots (Kumar {\em et al.}, 2000, and references therein). It has been 
generally accepted that the 
oscillatory power decreases in sunspots (Thomas, 1984; Braun, and Duvall, 1990; 
Lites {\em et al.}, 1998) and has been attributed to a variety of mechanisms, 
e.g. reduction in the efficiency of {\it p}-mode excitation by turbulent 
convection (Goldreich, and Keeley, 1977; Goldreich, and Kumar, 1988, 1990), absorption 
of acoustic power (Braun, Duvall, and LaBonte, 1988; Braun, 1995; Cally, 1995; 
Rosenthal and Julien, 2000), and modification of {\it p}-mode eigenfunctions by the 
magnetic field (Brown, 1994; Hindman, Jain, and Zweibel, 1997). The earlier 
study of Kumar {\em et al.}, (2000) showed a displacement in the peak of the 
acoustic sepctrum in sunspots to lower frequencies. 
Venkatakrishnan, Kumar, and Tripathy (2001) showed that the peak 
of the spectrum for small portions of the solar surface varies randomly with a spread 
of 200 $\mu$Hz resulting from the stochastic nature of the excitation. When several 
spectra were averaged, there was no discernible change in the peak within magnetic 
regions, and only a marginal N-S difference was detected between the hemispheres.

In this paper, we investigate the behaviour of the acoustic spectrum as a function of 
the underlying magnetic field using GONG+ data, which has a spatial scale of 
$\sim$ 2 arc-sec per pixel. We find that the high frequency 
power increases with magnetic field while the low frequency power decreases, with a 
transition at $\simeq$ 5 mHz. This corresponds to the predicted 
acoustic cut-off frequency of 
about 5.3 mHz at the temperature minimum for the quiet Sun in theoretical approaches 
to model the solar atmosphere (Gurman, and Leibacher, 1984; Balmforth and Gough, 1990). 
Observationally, some estimates for the quiet Sun have already been obtained 
(Fossat {\em et al.}, 1992, and references therein), yielding results which lie 
between 5.3 and 5.7 mHz. 
Hindman and Brown (1998) have found a similar increase in high frequency powers 
using SOHO/MDI data. In section~2, we describe our analysis and present our 
results, while these results are compared with the SOHO/MDI results in section~3.

\input epsf
\begin{figure}
\begin{center}
\leavevmode
\epsfxsize=4.5in\epsfbox{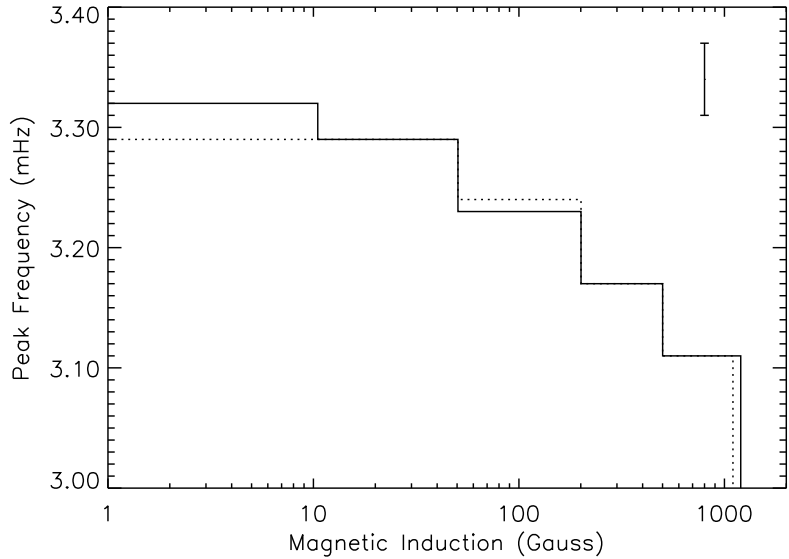}
\caption{Histograms representing the peak frequency of average power spectra over pixels having 
different values of magnetic inductions for 11 May, 2000 ({\it dashed line}) and 
12 June, 2000 ({\it solid line}). The error-bar at the right top corner 
indicates $1\sigma$ value.} 
\end{center}
\end{figure}

\input epsf
\begin{figure}
\begin{center}
\leavevmode
\epsfxsize=4.5in\epsfbox{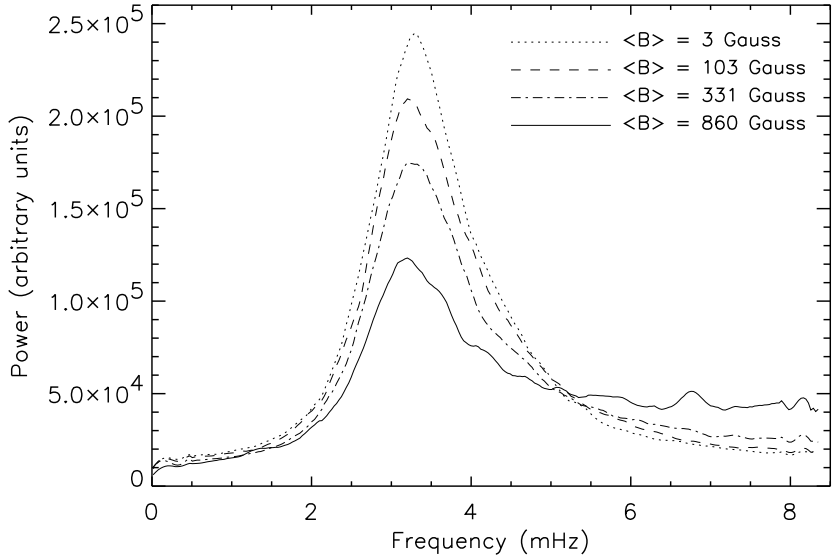}
\caption{Plots showing the average power spectra over pixels having different values of 
mean magnetic inductions, which are obtained after averaging on both the days. 
The average magnetic induction over each set of pixels is indicated in the figure.} 
\end{center}
\end{figure}

\input epsf
\begin{figure}
\begin{center}
\leavevmode
\epsfxsize=4.5in\epsfbox{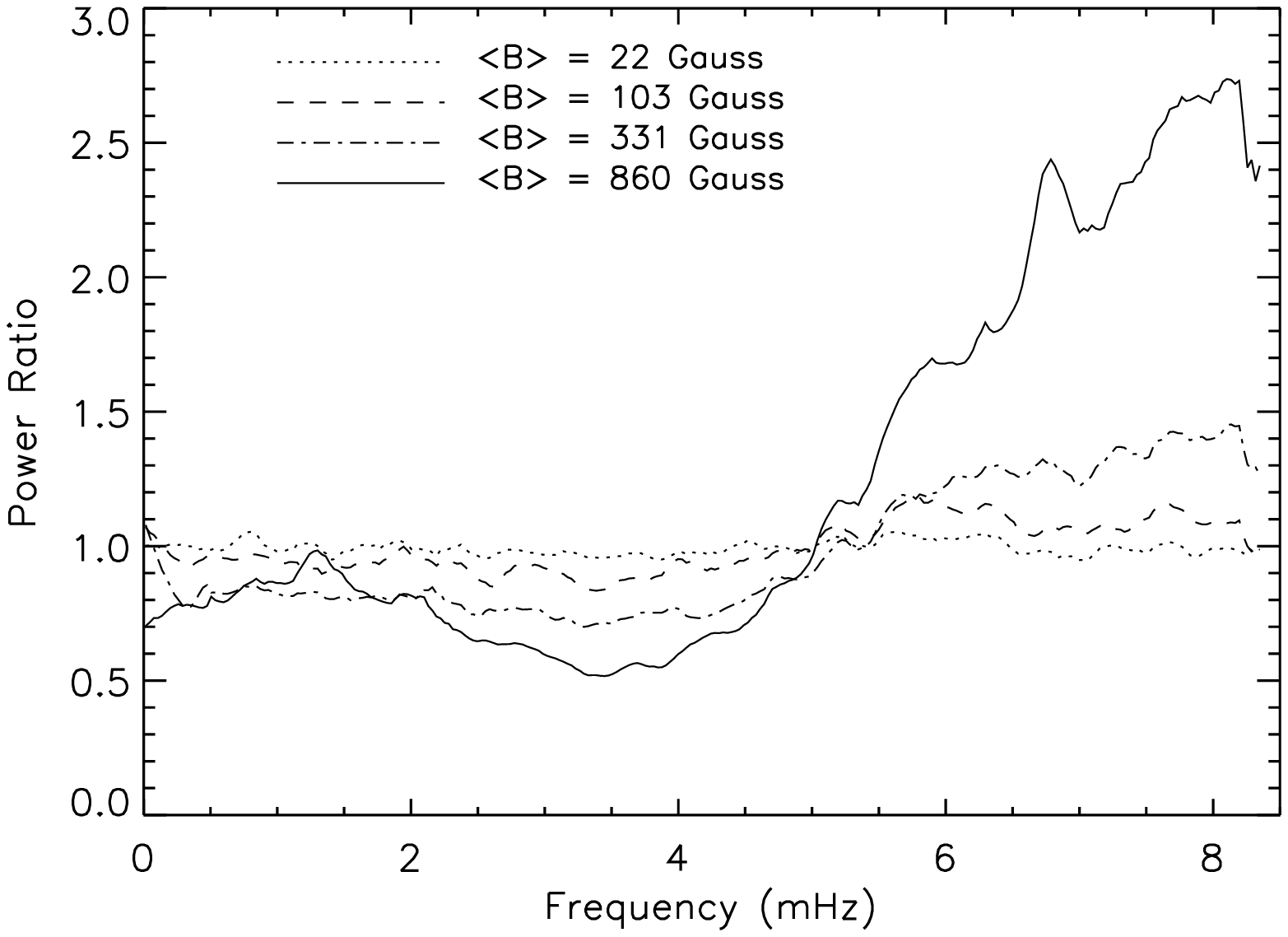}
\caption{Plots showing the ratio of powers in magnetized regions to that in quiet regions, which 
are obtained after averaging on both the days. The average magnetic induction over 
each set of pixels is indicated in the figure} 
\end{center}
\end{figure}

\section{Analysis and Results}
The data used here are from a time series of Dopplergrams for 11 May, 2000 and 
12 June, 2000
for 516 minutes duration starting at 15:00 UT obtained by the
GONG+ bread-board instrument at Tucson with a cadence of
1 minute. The Dopplergrams were corrected for mean solar rotation and
registered with the first Dopplergram of each day. It is well known that 
the Dopplergrams, in addition to the {\it p}-modes, also exhibit a variety of 
features, such as supergranulation pattern, meridional flows and solar rotation 
gradients. A two-point backward
difference filter was applied to the Dopplergrams to reduce low frequency drifts. 
The temporal power spectrum was
obtained for each pixel within the latitude and longitude range 
of $50^{o}$E $-$ $50^{o}$W 
and $50^{o}$N $-$ $50^{o}$S and then binned according to magnetic field strengths: 
0-10 Gauss, 10-50 Gauss, 50-200 Gauss, 200-500 Gauss, 500 Gauss \& above. A 
mean power spectrum was constructed for each magnetic field range. 
Further, the mean spectrum was smoothed by applying a digital filter, 
namely a Savitzky-Golay (S-G) filter (Press {\em et al.}, 1992). 
This filter basically smooths the data by a window 
function of a predefined number of data points and a polynomial order with a proper 
weighting and then finds the maximum of the smoothed spectrum. We observe the 
best fit with a window of 32 data points and a polynomial of 
order 6. The peak frequency shows a slight, but consistent decrease with increasing 
magnetic field for both the days (Figure~1). We find a maximum decrease of 
200$\pm$60 $\mu$Hz in peak frequency.

Figure~2 shows average power spectra over pixels having magnetic induction 
values in the ranges: 0-10 Gauss, 50-200 Gauss, 200-500 Gauss and 500 Gauss \& above. 
These spectra are obtained after averaging on both the days, 11 May, 2000 and 12 June, 2000. 
It is observed that the spectra intersect 
at a common point at $\simeq$ 5 mHz. In order to compare our results with 
those of Hindman and Brown (1998), we have also plotted the ratio of powers in 
magnetized regions to that in quiet regions in Figure~3.

\section{Discussions}
It has been already noted that high frequency oscillations like 3 minute 
oscillations are enhanced 
in sunspots (Beckers and Schultz, 1972; Kneer and Uexkuell, 1983; 
Thomas {\em et al.}, 1987; Horn, Staude, and Landgraf, 1997). 
Our results show that there is a general increase in the high frequency power. 
What is more significant is the fact that this enhancement is a function of the magnetic 
field. The most remarkable fact is the transition of behaviour from low frequency to high 
frequency around 5 mHz corresponding to the acoustic cut-off frequency of the quiet photosphere 
at temperature minimum. Our results are similar to those of Hindman and Brown (1998) 
at different frequencies and magnetic fields. The major difference seems to 
be a deficiency in power at 6 mHz as compared to Hindman and Brown (1998). For other 
frequency bands, even the numerical values of the power ratios are 
equal to the SOHO/MDI results. This augurs well for the fully deployed GONG+ system 
and establishes the potential of the new GONG instrument for the work in local 
helioseismology. A large number of active regions have to be examined before one can conclude 
on the deficiency in power at 6 mHz relative to the Hindman and Brown (1998) results. 

In summary, it can be noted that both GONG+ and SOHO/MDI data show that the ratio of 
acoustic power in magnetic regions to that in the quiet Sun is a function of 
the magnetic field over a large range of frequencies.

\acknowledgements
This work utilizes data obtained by the Global Oscillation Network
Group (GONG) project, managed by the National Solar Observatory, which is
operated by AURA, Inc. under a cooperative agreement with the National
Science Foundation. We deeply appreciate the advice provided by 
C. Lindsey. We are also thankful to anonymous referee for the suggestions which 
helped in improving the manuscript.

\end{article}
\end{document}